\documentclass[%
 reprint,
superscriptaddress,
 amsmath,amssymb,
 aps,
pra,
]{revtex4-2}

\usepackage{graphicx}% Include figure files
\usepackage{dcolumn}% Align table columns on decimal point
\usepackage{bm}% bold math
\usepackage{physics}
\usepackage{float}
\usepackage{multirow}
\usepackage{gensymb}
\usepackage{svg}
\usepackage{hyperref}% add hypertext capabilities
\usepackage[mathlines]{lineno}% Enable numbering of text and display math
\usepackage{comment}
\emergencystretch=\maxdimen
\begin{document}

\preprint{APS/123-QED}

\title{Full-field mapping of spatially varying polarization entanglement generated from spontaneous parametric down-conversion}

\author{Cheng Li}
\email{cli7@lbl.gov}
\altaffiliation[Present address: ]{Lawrence Berkeley National Laboratory, Berkeley, California 94720, USA}
\affiliation{Department of Physics, University of Ottawa, Ottawa, ON K1N 6N5, Canada}

\author{Girish Kulkarni}
\affiliation{Department of Physics, University of Ottawa, Ottawa, ON K1N 6N5, Canada}
\affiliation{Department of Physics, Indian Institute of Technology Ropar, Rupnagar, Punjab 140001, India.}

\author{Isaac Soward}
\affiliation{Department of Physics, University of Ottawa, Ottawa, ON K1N 6N5, Canada}

\author{Yingwen Zhang}
\affiliation{Department of Physics, University of Ottawa, Ottawa, ON K1N 6N5, Canada}
\affiliation{National Research Council, Ottawa, ON K1A 0R6, Canada}

\author{Jeremy Upham}
\affiliation{Department of Physics, University of Ottawa, Ottawa, ON K1N 6N5, Canada}

\author{Duncan England}
\affiliation{National Research Council, Ottawa, ON K1A 0R6, Canada}

\author{Andrei Nomerotski}
\affiliation{Faculty of Nuclear Sciences and Physical Engineering, Czech Technical University, Prague 115 19, Czech Republic}
\affiliation{Department of Electrical and Computer Engineering, Florida International University, Miami, FL 33174, USA}

\author{Ebrahim Karimi}
\affiliation{Department of Physics, University of Ottawa, Ottawa, ON K1N 6N5, Canada}
\affiliation{National Research Council, Ottawa, ON K1A 0R6, Canada}
\affiliation{Institute for Quantum Studies, Chapman University, Orange, CA 92866, USA}

\author{Robert Boyd}
\email{rboyd@uottawa.ca}
\affiliation{Department of Physics, University of Ottawa, Ottawa, ON K1N 6N5, Canada}
\affiliation{Institute of Optics, University of Rochester, Rochester, NY 14627, USA}


\begin{abstract}
Two-photon states generated from spontaneous parametric down-conversion (SPDC) can display entanglement in all degrees of freedom (DoFs) of light, including spatial, temporal, and polarization. The coupling between different DoFs of a two-photon state has been shown to display rich structures that enable novel and robust information processing schemes. While existing literature has studied these couplings by post-selecting the SPDC field, a comprehensive understanding of the inherent spatial-polarization coupling produced in the SPDC process is still lacking. This work produces a full spatial map of the polarization entanglement generated across the entire SPDC field, which contains an entire class of near-maximally polarization-entangled states with an average concurrence of $0.8303\pm0.0004$. The spatial-polarization coupling manifests as radially or linearly varying polarization-entangled states, whose wavefunctions are dependent on the transverse momenta of the down-converted photons and the pump beam, respectively. Our study lays important groundwork for further exploiting the coupling between entanglement in different DoFs for future quantum technologies.
\end{abstract}

\maketitle

%\tableofcontents

%\section{\label{sec:level1}Introduction}

\section{Introduction}\label{sec1}

Spontaneous parametric down-conversion (SPDC) has been a major workhorse for generating entangled tw-photon states \cite{Harris1967prl,burnham1970PRL,Shih1988PRL}, enabling numerous applications in quantum information processing. In SPDC, a photon from a pump beam, which has a higher optical frequency, interacts with a second-order nonlinear medium to be annihilated and produce a pair of down-converted photons with lower frequencies \cite{burnham1970PRL}. The down-converted photons, which are commonly referred to as signal and idler, can exhibit entanglement in all degrees of freedom (DoFs) of light, including spatial, temporal, and polarization. 

Studying the coupling between different DoFs is a crucial step towards a comprehensive understanding of SPDC-based entanglement generation and engineering novel entangled photon states for practical applications \cite{he2022lsa,nape2023aplphot}. For instance, earlier studies have shown that the coupling between spatial and temporal-spectral DoFs can be used to tailor the entanglement of quantum states generated from SPDC \cite{Gatti2009prl, Baghdasaryan2022pra, Sevilla-Gutierrez2024pra}. An equally important aspect of this subject is the coupling between spatial and polarization DoFs in SPDC. Such cross-DoF couplings can be exploited to enable novel and efficient quantum information applications. For instance, it was shown that quantum correlations across spatial and polarization DoFs can enable novel and more robust information encoding schemes and demonstrate topological resilience against environmental noise \cite{Ornelas2024NP}. Moreover, the spatially varying polarization correlations in a photonic entangled state can also be harnessed to enable quantum holographic imaging \cite{defienne2021natphys} or engineer photonic cluster states for efficient quantum computation \cite{Vallone2010PRA, Ciampini2016Light}. From a fundamental perspective, studies have shown that spatial modes of the pump beam can influence the polarization entanglement of the down-converted photons \cite{li2023PRA, zhang2023PRApplied}, indicating the opportunity of directly engineering the couplings between different DoFs in a photonic entangled state, or encoding phase images in the two-photon correlations \cite{Verniere2024prl} by shaping the pump beam.

The spatial-polarization coupling in SPDC commonly manifests as two-photon states with spatially varying polarization entanglement. Although one can readily produce these features from SPDC using a paired crystal configuration \cite{kwiat1999pra, barreiro2005prl}, the large spatial dimensionality of the SPDC field hinders the efficient characterization of them. For instance, a full state tomography for a bipartite state with local dimensionality $d$ in a single DoF requires $\mathcal{O}(d^4)$ single-outcome projective measurements \cite{james2001PRA, friis2019natrevphys, HerreraValencia2020Quantum}. Since the total dimensionality of a quantum state scales with the product of dimensionality in each constituent DoF, the need to address correlations, both within and between different DoFs, only adds to the complexity of the problem. Without comprehensively resolving the polarization entanglement produced in each spatial mode, the rich structures resulting from the coupling between different DoFs reduce to a mixture of distinguishable states. As a result, the full potential of SPDC-induced spatial-polarization coupling becomes inaccessible, limiting its application in practical quantum devices. For instance, a recent work indicates that even though each spatial mode in a multimode pump beam can generate maximum polarization entanglement through SPDC, detecting the two-photon states without resolving individual spatial modes can make the entanglement appear deteriorated \cite{li2023PRA}.

\begin{figure*}[ht]%
\centering
\includegraphics[width=0.8\textwidth]{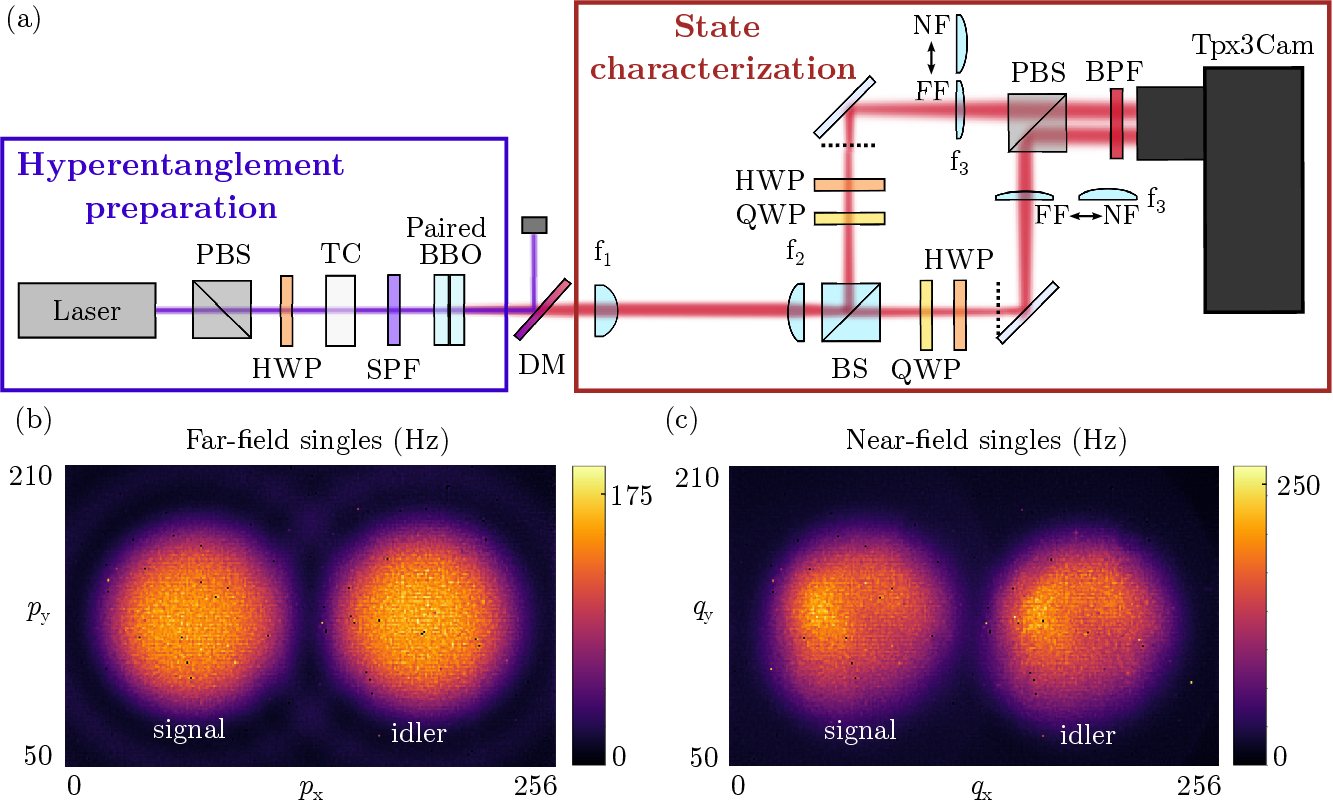}
\caption{(a) Schematic of the experimental setup. The $\beta$-barium borate (BBO) double-crystal produces a two-photon state, which exhibits spatially varying polarization entanglement. The Tpx3Cam captures the two-photon field in the momentum or position basis through different lens configurations and in different polarization bases through combinations of the quarter-wave plate (QWP), half-wave plate (HWP), and polarizing beam splitter (PBS). TC: temporal compensator, SPF: short-pass filter with cutoff wavelength at 500 nm, DM: dichroic mirror, BS: beam splitter, BPF: band-pass filter centered at 800 nm with a bandwidth of 40 nm. \(f_{1}\) = 50 mm, \(f_{2}\) = 100 mm. \(f_{3}\) = 150 mm for far-field measurements or 75 mm for near-field measurements. Dashed lines represent the intermediate planes imaging the near-field of the crystal using \(f_{1}\) and \(f_{2}\). (b-c) Time-stamp histogram of photons detected in the far-field and near-field of the crystal, representing projection onto the momentum (\textit{p}) and position (\textit{q}) bases of the two-photon state.}\label{fig1}
\end{figure*}

Over the past decade, several studies have employed Tpx3Cam, a data-driven camera capable of time-stamping single photons \cite{Fisher2016joi,nomerotski2019nuclear,Nomerotski2023joi}, to enable rapid characterization of photonic entangled states produced from SPDC \cite{ianzano2020SciRep,courme2023optlett,Gao2024PRL}. However, these works have not fully explored the coupling between different DoFs. Specifically, Ref.~\cite{ianzano2020SciRep, courme2023optlett} have focused only on the polarization DoF or the spatial DoF, respectively, while their photon pair sources are capable of producing spatially varying polarization-entangled states. As a result, the potential of the embedded spatial-polarization coupling has not been addressed. Furthermore, the entangled states produced in Ref.~\cite{ianzano2020SciRep, Gao2024PRL} are first spatially postselected using single-mode fibers or polarization-maintaining fibers before being imaged onto the Tpx3Cam, and any spatial structures in the resulting polarization are imprinted by external modulation devices. Doing so effectively reduces the spatial dimensionality and obscures any spatial-polarization coupling imposed by the generation process.  

In this work, we employ Tpx3Cam to comprehensively characterize the spatially varying polarization entanglement that is natively produced from SPDC. In contrast to earlier works, we image the full spatial field of SPDC without any post-processing on the resulting two-photon states prior to their detection. By performing spatially-resolved polarization state tomography, we confirm the generation of near-maximally polarization-entangled states across the entire spatial profile of the SPDC field, with an average concurrence of $0.8303\pm0.0004$. Importantly, we present a complete spatial map of the polarization state generated from SPDC, revealing how polarization entanglement depends on the transverse momenta of the down-converted photons throughout the entire spatial profile of the SPDC field. Moreover, using a weakly focused pump beam, we experimentally characterize the dependence of polarization entanglement on the angular spectrum of the pump beam. Our results advance the fundamental understanding of the interplay between different DoFs of a photonic entangled state, paving the way for novel applications in photonic quantum information processing.

\section{Experimental setup}\label{sec2}

Fig.~\ref{fig1}(a) depicts our experimental setup. An ultraviolet continuous-wave laser emits the pump beam with a center wavelength of 405 nm, a bandwidth of 2 nm, and a power of 20 mW. The pump beam is made polarized at $45^{\circ}$ using a polarizing beam splitter (PBS) and a half-wave plate (HWP). A 5-mm temporal compensator (TC) quartz crystal introduces a time delay between the horizontal (H-) and vertical (V-) components of the pump beam, which pre-compensates for the temporal walk-off that the two orthogonal polarization components are expected to subsequently gather inside the BBO double-crystal \cite{nambu2002pra}. The BBO crystals are each 0.5-mm-thick and identically cut for type-I phase-matching, with their optical axes oriented perpendicularly to one another \cite{kwiat1999pra}. A $45^{\circ}$-polarized photon from the pump beam can then be down-converted in either crystal with equal probability to produce a pair of H-polarized or V-polarized photons. In the experiment, we orient the double-crystal for near-collinear phase matching. The down-converted photons are separated from the pump beam using a dichroic mirror (DM), and then probabilistically split into two arms using a non-polarizing beam splitter (BS). We ignore the cases wherein both the down-converted photons exit the same port of the BS, and denote photons that are reflected as \textit{signal} and those that are transmitted as \textit{idler}. This beam-splitting scheme ensures that measuring the correlation between the signal arm and the idler arm resolves the full spatial profile of the SPDC. Although the probabilistic splitting reduces the coincidence rates by $50\%$, it does not affect the quality of entanglement or the measured spatial-polarization structure compared to those obtained with deterministic splitting, as can be seen later in Fig.~\ref{fig3} and Fig.~\ref{fig4}. In the low-gain regime, the output state $|\Psi\rangle$ in the far-field can be written in the joint spatial-polarization basis as
\begin{widetext}
\begin{equation}\label{eqn2.1.1}
    |\Psi\rangle = \frac{1}{\sqrt{2}}\sum_{\textbf{\textit{p}}_s,\textbf{\textit{p}}_i}c_{\textbf{\textit{p}}_s,\textbf{\textit{p}}_i}\bigg[|H,\textbf{\textit{p}}_s,H,\textbf{\textit{p}}_i\rangle+\mathrm{e}^{i\phi(\textbf{\textit{p}}_s,\textbf{\textit{p}}_i)}|V,\textbf{\textit{p}}_s,V,\textbf{\textit{p}}_i\rangle\bigg],
\end{equation}
\end{widetext}
where $\textbf{\textit{p}}_{s(i)}$ represents the transverse momentum of the signal (idler) photon and $c_{\textbf{\textit{p}}_s,\textbf{\textit{p}}_i}$ are complex coefficients. The quantity $\phi(\textbf{\textit{p}}_s,\textbf{\textit{p}}_i)$ represents a phase difference between the orthogonal polarization components of the state, which is dependent on the transverse momentum of the signal and the idler photons \cite{altepeter2005opex}. In what follows, we refer to $\phi$ as the \textit{two-photon polarization phase}. For two-photon states generated from paired Type-I SPDC crystals, given that $\textbf{\textit{p}}_{s}=-\textbf{\textit{p}}_{i}$, $\phi$ displays a quadratic dependence on the emission angle of the signal and idler photons \cite{Brida2007pra}, which can be characterized $|\textbf{\textit{p}}_{s(i)}|$. We then have
\begin{equation}\label{eqn2.1.2}
\phi(\textbf{\textit{p}}_s,\textbf{\textit{p}}_i) = k |\textbf{\textit{p}}_{s(i)}|^2 L + \phi_0,
\end{equation}
where $k$ is the wavevector of the SPDC photons, L is the length of each of the paired crystals, and $\phi_0$ is an overall phase between the SPDC processes in the two crystals, which can be influenced by the crystal length and cut angle.

The signal and idler photons then propagate through a quarter-wave plate (QWP) and a HWP before they are either transmitted or reflected by a PBS. The photons finally pass through a bandpass filter with a center wavelength of 800 nm and full width at half maximum (FWHM) bandwidth of 40 nm before being detected by the Tpx3Cam. Since the two-photon polarization phase has insignificant wavelength dependence around 810 nm, we have chosen a wider bandwidth for our spectral filter to increase the overall photon brightness without degrading the polarization entanglement. To characterize entanglement in the polarization DoF, we measure the polarization correlation between each pair of momentum-correlated signal and idler photons. We first image the far-field of the crystal onto the Tpx3Cam using the combination of $f_1=50$ mm, $f_2=100$ mm, and $f_3=150$ mm lenses, and then acquire data for 16 different combinations of orientations of HWPs and QWPs, each for 1 minute. These polarization measurements allow us to reconstruct the polarization density matrices corresponding to all pairs of momentum-correlated photons through quantum state tomography, which characterizes polarization entanglement. To characterize the state dimensionality in the spatial DoF, we measure the spatial correlation of signal and idler photons in the position basis. We do so by replacing the $f_3=150$ mm lenses with $f_{3}=75$ mm ones, thereby imaging the crystal output face (near-field) onto the Tpx3Cam sensor. We then set the HWPs and QWPs to project the polarization in both arms into V and acquire data for 1 minute. Combined with the 16 measurements acquired in polarization characterization, the total required data acquisition time is thus $1 + 16 = 17$ minutes. 

The imaging sensor in Tpx3Cam comprises a $256\times256$ pixel array with a pixel pitch of 55 $\mu$m. The pixels in Tpx3Cam are data-driven and individually trigger the registration of photon incidence events when the signal amplitude exceeds a predefined threshold. The camera can be single-photon sensitive with the addition of an image intensifier (Photonis Cricket) and has a single photon temporal resolution of 2 ns \cite{Nomerotski2023joi}. Time-stamping for individual photons can potentially allow for more versatility in data acquisition and analysis compared to frame-based imaging devices. During data acquisition, each incident photon could hit a cluster of pixels due to being amplified by the intensifier. To correct this, we apply a centroiding algorithm that identifies the amplitude-weighed center in each cluster as the true pixel coordinate. We use the time stamp of each centroided pixel as a reference to correct the time walk within the cluster. After centroiding and time walk correction, we apply a two-pointer technique to the sorted time stamps of signal and idler photons and identify events detected within a 10-ns time window as photon coincidences \cite{ianzano2020SciRep}. Although the time walk correction algorithm can reduce the temporal resolution to 8 ns \cite{Vidyapin2023SciRep, Zhang2020PRA}, it should not significantly affect the efficiency of the subsequent coincidence counting with our choice of a longer time window. As a result, we obtain an average coincidence rate of approximately 2.7 Hz between two spatially correlated $3\times3$-pixel regions, which have similar sizes with the correlation widths measured at the Tpx3Cam sensor plane (see Section~\ref{sec3.1} for details). Throughout this paper, we refer to each $3\times3$-pixel region as a superpixel. Moreover, we observe an overall coincidence rate of 3200 Hz between the full signal and idler fields. We note that although the experimentally measured coincidence rate is limited by the $8\%$ overall photon detection efficiency of the Tpx3Cam system \cite{Vidyapin2023SciRep}, our choice of 1-minute data acquisition time has allowed us to obtain enough counts for the relevant results to be statistically valid.

\section{Results and discussions}\label{sec3}

\subsection{Characterization of spatial correlations}\label{sec3.1}
\begin{figure}[ht]%
\centering
\includegraphics[width=0.48\textwidth]{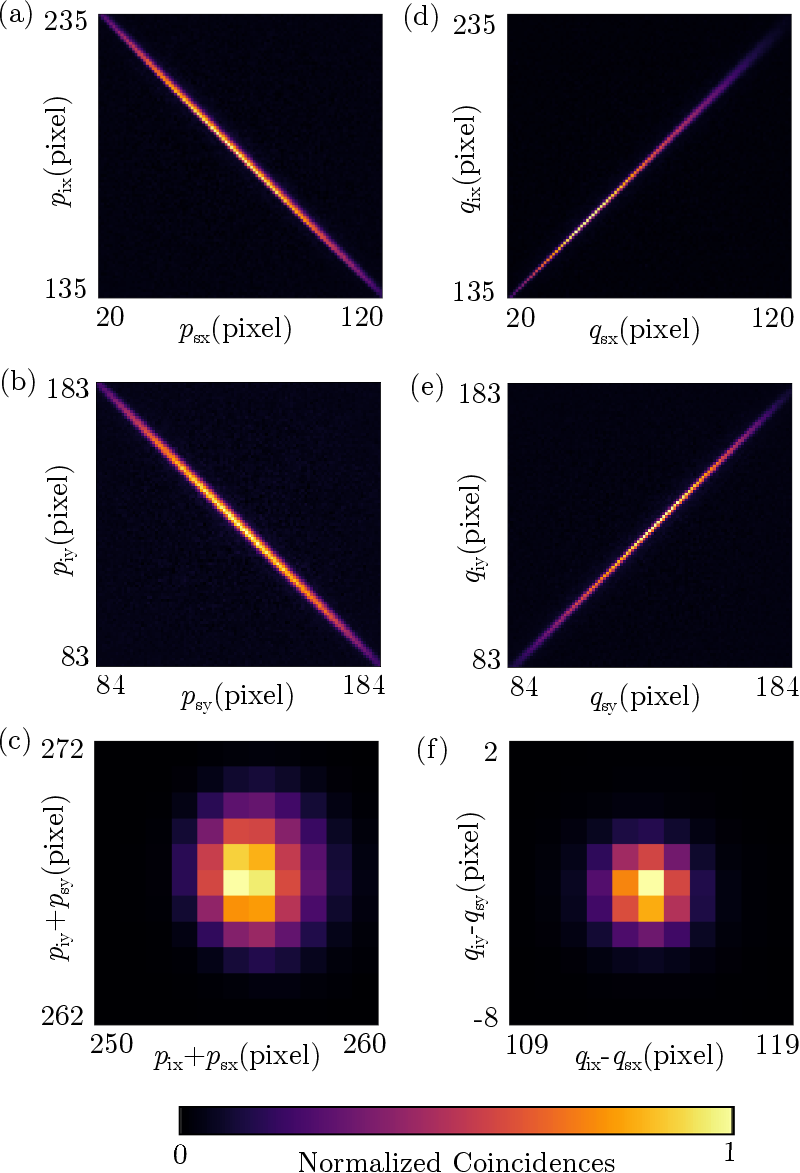}
\caption{Spatial correlation profiles measured in the $x$-components, $y$-components, and joint coordinates of the (a-c) momentum and (d-f) position of the signal-idler photons.}\label{fig2}
\end{figure}

In Fig.~\ref{fig2}, we depict the experimentally measured $x$- and $y$-components of the spatial correlation profile in both the position and momentum basis. We note the strong position correlations and momentum anti-correlations between the photons, which is a characteristic feature of spatial entanglement. Specifically, Fig.~\ref{fig2}(c) and (f) depict the biphoton correlation profiles in the joint momentum and position coordinates of the down-converted photons. From these results, we conclude that in the Tpx3Cam sensor plane, individual discrete momentum modes are best approximated by $3\times3$-pixel large superpixels. For this reason, we characterize the polarization entanglement between each pair of correlated superpixels in the far-field. 

The results also allow us to quantitatively verify spatial entanglement by demonstrating violations of the Einstein-Podolsky-Rosen (EPR) criteria \cite{reid1989pra, bhattacharjee2022njp}. We first obtain near-field (NF) and far-field (FF) correlation widths $\Delta_\text{NF}$ and $\Delta_\text{FF}$, respectively, at the Tpx3Cam sensor plane by fitting Gaussians to the spatial correlation profiles of Fig.~\ref{fig2} in the $x$- and $y$-directions \cite{monken1998pra, law2004prl, bhattacharjee2022njp}. We then calculate the position and momentum uncertainties using the relations
\begin{subequations}
 \begin{equation}\label{eqn2.1.3}
    \Delta(p_{ir}|p_{sr}) = \frac{k_{si}\hbar}{f_e}\Delta_\text{FF},
\end{equation}
\begin{equation}\label{eqn2.1.4}
    \Delta(q_{ir}|q_{sr}) = \frac{1}{M}\Delta_\text{NF},
\end{equation}
\end{subequations}
where $p_{sr(ir)}$ and $q_{sr(ir)}$ stand for the momentum and position of the signal (idler) photons, respectively, $r = x, y$ represent the x- and y-components of the quantities, $k_{si} = (2\pi/810)$ nm$^{-1}$ is the wavevector of the signal and idler photons, $f_e = 75$ mm is the effective focal length of our far-field imaging system and $M=2$ is the magnification of our near-field imaging system. We summarize our results in Table 1 and compute the conditional Heisenberg uncertainty products as
\begin{subequations}
\begin{equation}\label{eqn2.1.5}
    \Delta_\text{min} p_{x} \Delta_\text{min} q_{x} = (0.11 \pm 0.05) \hbar < \hbar/2,
\end{equation}
\begin{equation}\label{eqn2.1.6}
    \Delta_\text{min} p_{y} \Delta_\text{min} q_{y} = (0.12 \pm 0.03) \hbar < \hbar/2,
\end{equation}
\end{subequations}
which clearly violates the EPR criteria in both the x- and y-directions. This indicates that the two-photon state is spatially highly entangled.

The spatial dimensionality of the prepared quantum state can be estimated from the ratio $\Delta \textbf{\textit{p}}_+/\Delta \textbf{\textit{p}}_-$, where $\Delta \textbf{\textit{p}}_{\pm}$ stands for the joint measurement uncertainty of $\textbf{\textit{p}}_s\pm\textbf{\textit{p}}_i$. The values of $\Delta \textbf{\textit{p}}_{\pm}$ can be inferred by performing Gaussian fitting along the diagonal and anti-diagonal directions of JPD results measured in the momentum basis. As a result, the entangled two-photon state is estimated to contain 2496 spatial modes in the momentum basis, which implies high dimensionality in the spatial DoF.

We note that the asymmetry observed in the x-components of the position correlation, which is depicted in Fig.~\ref{fig2}(d), is likely a consequence of a distortion in the pump's intensity profile. As depicted in Fig. 1(c) in the main text, the near-field singles rates display a non-Gaussian spatial profile as a result of the distorted pump profile. Although this imperfection may affect the verification of spatial entanglement using the EPR criterion, which requires the pump beam to have a perfectly Gaussian profile \cite{law2004prl}, it does not refute our conclusions regarding high entanglement dimensionality in the spatial DoF. Nevertheless, to certify entanglement dimensionality, one needs to adopt an assumption-free protocol, such as estimating a lower bound of Schmidt number by measuring correlations in two mutually unbiased bases \cite{erker2017quantum, bavaresco2018natphys, ndagano2020npjqi}. We may explore this direction in a future work.

\begin{table}[htbp]
\caption{Measurement uncertainties inferred from spatial correlation profiles}\label{tab1}%
\centering
\begin{tabular}{@{}llll@{}}
\hline
Quantity & Values & Units\\
\hline
$\Delta_\text{min} p_{x}$   & $(4.9 \pm 0.2) \times 10^{-3}$  & $\hbar/\mu$m\\
$\Delta_\text{min} p_{y}$    & $(6.4 \pm 0.3) \times 10^{-3}$  & $\hbar/\mu$m\\
$\Delta_\text{min} q_{x}$    & $18.76 \pm 9.49$  & $\mu$m\\
$\Delta_\text{min} q_{y}$    & $18.18 \pm 3.63$  & $\mu$m\\
\hline
\end{tabular}
\end{table}

\subsection{Full spatial mapping of polarization entanglement}\label{sec3.2}
Using the setup depicted in Fig.~\ref{fig1}(a), we perform quantum state tomography \cite{james2001PRA} using experimentally measured polarization correlation between each pair of momentum-correlated (diametrically-opposite) superpixels in the far-field and reconstruct the corresponding polarization density matrix. We then compute the concurrence using $C = \text{max}\{0, \lambda_1-\lambda_2-\lambda_3-\lambda_4\}$, where $\lambda$'s are the eigenvalues of a Hermitian matrix derived from applying Pauli-y operations on reconstructed density matrices \cite{wootters1998prl}. The two-photon polarization phases $\phi$ are computed as the phase of the $|VV\rangle\langle HH|$ elements of the reconstructed density matrices.

\begin{figure*}[ht]%
\centering
\includegraphics[width=\textwidth]{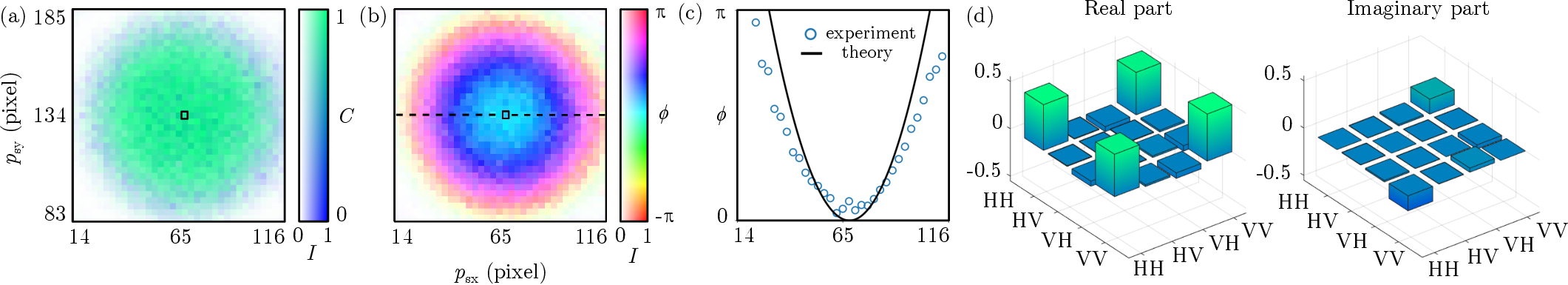}
\caption{(a) Concurrence $C$ and (b) two-photon polarization phase $\phi$ of the two-photon states measured between momentum-correlated superpixels. The pixel coordinates correspond to the central position of the superpixels for signal photons. In the colormap, the saturation depicts the normalized pair generation rate, $I$, and the hue depicts $C$ and $\phi$ in (a) and (b), respectively. (c) Comparison between of the theoretical description and experimental measurement of $\phi$ along the dashed line in (b). (d) Real and imaginary parts of the density matrix reconstructed at the center of the SPDC field, the corresponding signal pixel is marked with a black square in (a) and (b).}\label{fig3}
\end{figure*}

Fig.~\ref{fig3} depicts the full spatial maps of the concurrence $C$ and two-photon polarization phase $\phi$ in the far field. For illustration purposes, we have mapped all signal-idler pairs to the central coordinates of the superpixels of signal photons. In Fig.~\ref{fig3}(d), we also display a two-photon polarization density matrix, which is reconstructed at the center of the SPDC field, with the corresponding signal-idler pair indicated by a black square in Fig.~\ref{fig3}(a-b). We calculate an average concurrence of 0.8303 $\pm$ 0.0004, which implies strong polarization entanglement with a spatially varying structure across the entire SPDC field. Here, we have defined the average concurrence as the mean value of local concurrences calculated for individual momentum-correlated superpixel pairs. Since the $\phi$ does not influence concurrence, this definition allows us to characterize the expected polarization entanglement across the entire SPDC field regardless of the phase structure. 

We observe from Fig.~\ref{fig3}(b) that the two-photon polarization phase $\phi$ displays a gradient along the radial direction. In Fig.~\ref{fig3}(c), we depict the dependence of $\phi$ on $\textbf{\textit{p}}_{s}$ along the x-direction and observe that the spatial distribution of $\phi$ agrees well with the prediction of Eq.~\ref{eqn2.1.2} and a partial spatial map reported in Ref.~\cite{altepeter2005opex}. The discrepancies towards the edge of the SPDC field can be attributed to transverse walk-offs due to propagation in a birefringent crystal. One can perform a similar measurement in the near field. However, we expect such measurements to yield low entanglement over all position-correlated superpixels with no well-defined phase structure. The reason is that detecting each pair in the position basis effectively mixes two-photon states with different $\phi$'s in the momentum basis, thereby lowering the detected polarization entanglement.

Due to the high spatial entanglement and polarization entanglement across the entire SPDC field, we also expect a high dimensionality over the joint spatial-polarization state. Although a rigorous estimation of such dimensionality requires measurements in the joint spatial-polarization bases, we can estimate a total number of spatial-polarization modes from the results presented in Fig.~\ref{fig2} and \ref{fig3} by making the following observation. We note that $\phi$ is highly uniform along the azimuthal direction, indicating highly indistinguishable polarization entanglement at each circular region concentric with the SPDC field. Consequently, if each corresponding spatial mode is also highly indistinguishable from one another, the two-photon state can be seen as being hyperentangled \cite{Kwiat1997jmo, barreiro2005prl}, which is a tensor product of (i) spatially entangled discrete momentum modes occupying the full spatial field of SPDC; and (ii) polarization-entangled states with a well-defined two-photon polarization phase $\phi$. In principle, $\phi$ can be compensated [for example,  using spatial light modulators (SLMs)] so that it becomes uniform also across all radial positions without reducing the polarization entanglement \cite{defienne2021natphys}. As a result, we expect the attainable total dimensionality of our state to be equal to that of a global hyperentangled state, which is a product of the average dimensionality in the polarization DoF and the estimated number of modes in the spatial DoF, as underscored by the tensor product postulate of quantum physics. 

Although two-photon polarization states reside in a two-dimensional Hilbert space, the effective dimensionality may be smaller for non-maximally entangled states. Therefore, it is pertinent to estimate a lower bound of the dimensionality $d$ in the polarization DoF using the relation $\log_2d \leq E$ \cite{Nielsen_Chuang_2010}, where $E$ stands for the entanglement of formation \cite{wootters2001qic}. For two-photon polarization states, the entanglement of formation can be derived from concurrence using \cite{wootters1998prl}
\begin{equation}
    E(C) = h\left(\frac{1 + \sqrt{1 - C^2}}{2}\right),
\end{equation}
where $h(x) = -x \log_2{x} - (1 - x) \log_2{(1 - x)}$. The average entanglement of formation is then estimated to be $E = 0.7695 \pm 0.0005$. Combing the total number of spatial modes estimated in Section.~\ref{sec3.1}, we estimate the total attainable dimensionality of the entangled state to be $2496\times2^E\approx4254$.

\subsection{Native spatial-polarization coupling in SPDC processes}

The spatial-polarization coupling in SPDC has two important aspects. First, the exact form of polarization entanglement depends on the transverse momenta of the \textit{down-converted photons}. This effect, which is displayed in Fig.~\ref{fig3}(b), occurs because down-converted photons emitted with different transverse momenta experience angle-dependent refractive indices inside the birefringent nonlinear medium and consequently accumulate different phase retardations between orthogonal polarization components. Consequently, a class of highly entangled states of the form of Eq.~(\ref{eqn2.1.1}) with spatially varying $\phi$ is generated across the spatial profile of the field. While the spatial profile of polarization entanglement in SPDC has been studied for limited spatial regions or specific sets of spatial modes \cite{barreiro2005prl,altepeter2005opex, Brida2007pra}, our results present the first full spatial distribution of polarization entanglement in the entire spatial profile of the field, laying the foundation for the controlled generation of desired photonic entangled states. 

Phase-matching of the crystal also controls the relation between two-photon transverse momenta and the two-photon polarization phase. As may be noted from Fig.~\ref{fig3}(b), the two-photon polarization phase $\phi$ does not vary over the full range of $-\pi$ to $\pi$ in the present configuration as a consequence of the near-collinear phase matching condition. In what follows, we tilt the crystal to slightly increase the angle between the pump beam's wavevector and the crystals' optics axes, thereby enlarging the emission angle of the SPDC field \cite{Karan_2020}. As dictated by Eq.~(\ref{eqn2.1.2}), $\phi$ varies faster with transverse momenta at larger emission angles, with a possibility of spanning over the full parameter space in the spatial profile of the field, which allows access to a much wider class of polarization-entangled states.

\begin{figure*}[ht]%
\centering
\includegraphics[width=0.8\textwidth]{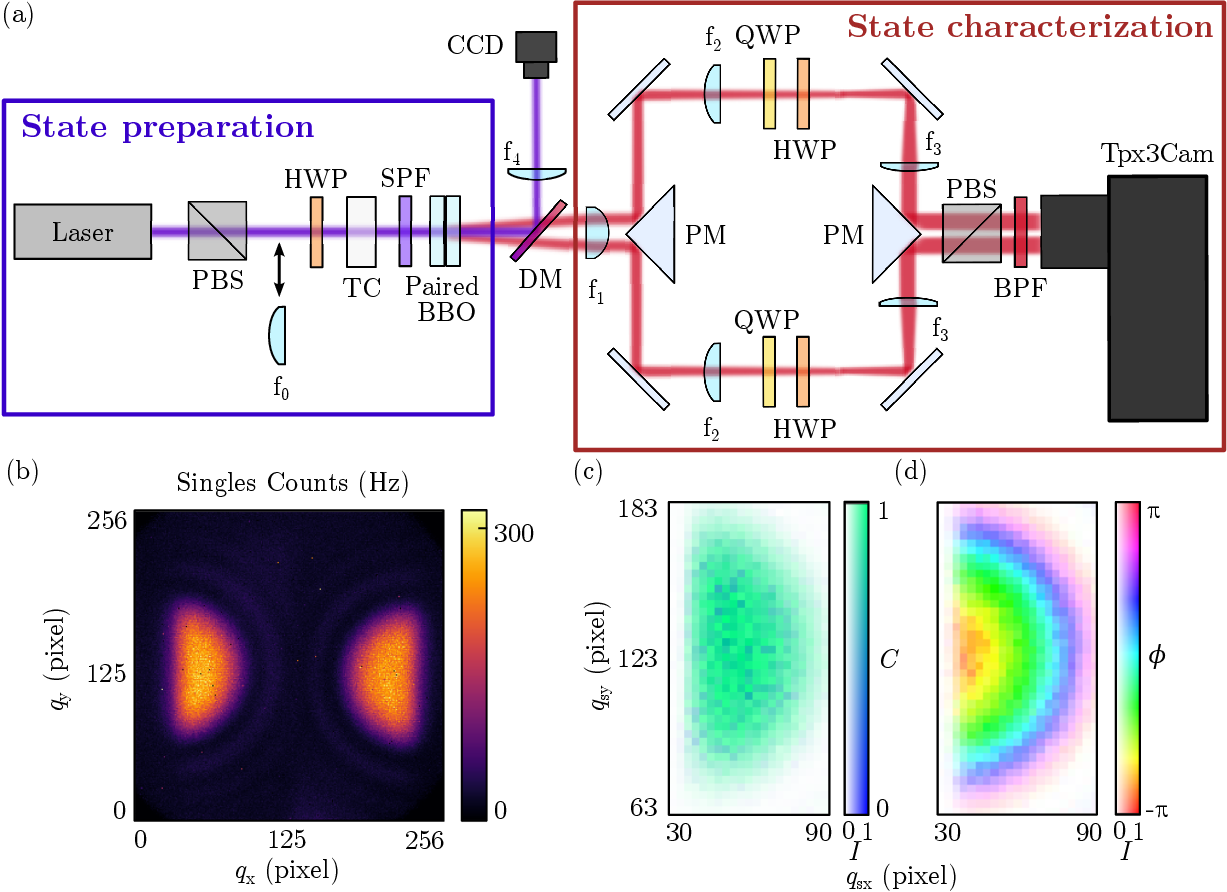}
\caption{(a) Schematic of the modified experimental setup. PM: prism mirror. Focal lengths of the lenses are \(f_{0}\) = 100 mm, \(f_{1}\) = 50 mm, \(f_{2}\) = 100 mm \(f_{3}\) = 150 mm and \(f_{4}\) = 75 mm. The angular width of the pump beam is changed by inserting \(f_{0}\) into the beam path. A CCD camera images the pump beam in the Fourier plane to analyze its angular width. (b) Far-field image of the SPDC field taken by the Tpx3Cam. (c) Concurrence and (d) relative phase of the two-photon states measured between momentum-correlated superpixels. Pixel coordinates correspond to the central position of the superpixels for signal photons. }\label{fig4}
\end{figure*}

We depict our modified setup in Fig.~\ref{fig4}(a). In order to capture the enlarged far-field image profile in its entirety, we deterministically split the SPDC field into two half-circles using a prism mirror (PM) and recombine the fields onto the Tpx3Cam sensor. We retain the same far-field imaging scheme, such that the two-photon correlation width remains unchanged from the earlier setup, and we again perform spatially resolved polarization state tomography using the same procedure as discussed previously. In Fig.~\ref{fig4}(b),(c), and (d), we depict the far-field intensity Tpx3Cam image, concurrence $C$, and two-photon polarization phase $\phi$, respectively. We again observe a strong polarization entanglement in the entire spatial extent of the field with an average concurrence of $0.8847\pm0.0006$. We notice a marginal increase in the average concurrence compared to the result in Sec.~\ref{sec3.2}. This is likely a result of the lower polarization cross-talk since now both signal and idler photons are detected through the transmission port of the PBS. The distribution of $\phi$ again displays a gradient along the radial direction, but this time with $\phi$ having an enlarged parameter space spanning from $-\pi$ to $\pi$. In other words, the setup produces near-maximally entangled states of Eq.~(\ref{eqn2.1.1}) with all possible $\phi$. Thus, our setup can be configured to supply any specific state with a desired value of $\phi$ by post-selecting the corresponding pair of far-field pixels and tuning the phase-matching of the crystal. Such a mechanism allows encoding spatial mode information in polarization correlations or vice versa, which could have important implications for quantum key distribution protocols using qudit-like states \cite{Scarfe2025commphy}. It is also possible to structure $\phi$ to have any desired profile by introducing additional phase differences between H- and V-polarized down-converted photons using SLMs to enable holographic quantum imaging \cite{defienne2021natphys} with much higher speed. We note that SLMs offer a higher degree of control over the polarization structure at the expense of significant insertion loss. Therefore, harnessing the native spatial-polarization coupling from the SPDC process could serve as a low-loss alternative for structuring quantum states.

\begin{figure*}[ht]%
\centering
\includegraphics[width=0.8\textwidth]{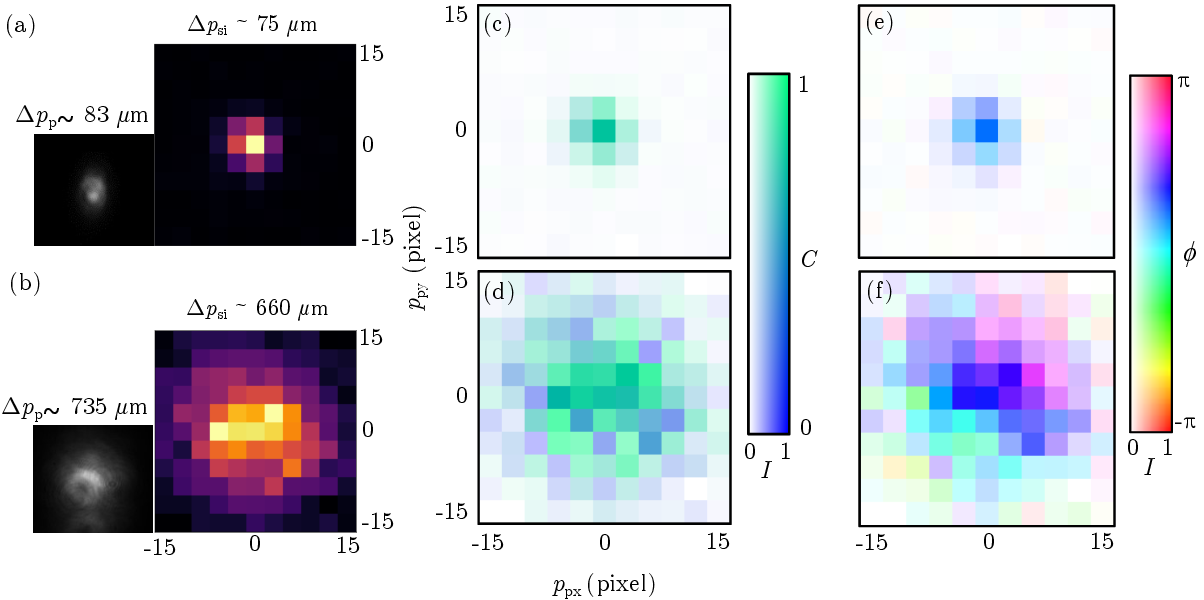}
\caption{Pump beam profile (greyscale image) and two-photon momentum correlation profile (colored image) for (a) collimated and (b) weakly focused pump beam. The two-photon momentum correlation profiles display the coincidences between different idler superpixels and the same signal superpixel, from which one can infer the influence of pump beam’s angular spectrum on the width of two-photon momentum correlation. The inset captions state the beam width measured at the camera sensor plane. (c-d) Concurrence and (e-f) the two-photon polarization phase measured between a fixed signal superpixel and all momentum-correlated idler superpixels, where (c) and (e) correspond to collimated pumped beam while (d) and (f) correspond to weakly focused pump beam. Pixel coordinates indicate the transverse momenta of the pump beam $\textbf{\textit{p}}_\text{p} = \textbf{\textit{p}}_s + \textbf{\textit{p}}_i$, in which the geometrical centers of two-photon momentum correlation profile are denoted $\textbf{\textit{p}}_\text{px} = \textbf{\textit{p}}_\text{py} = 0$} \label{fig5}
\end{figure*}

The second important aspect of spatial-polarization structure is between the polarization DoF of the down-converted photons and the spatial DoF of the \textit{pump beam}. Specifically, the two-photon polarization phase $\phi(\textbf{\textit{p}}_s,\textbf{\textit{p}}_i)$ in Eqn. (\ref{eqn2.1.1}) depends on the transverse momentum of the pump beam since $\textbf{\textit{p}}_\text{p} = \textbf{\textit{p}}_s + \textbf{\textit{p}}_i$ \cite{li2023PRA}. This relation is first theoretically quantified by Li et al. \cite{li2023PRA}. Here, to experimentally characterize this influence, we introduce a broader angular spectrum in the pump beam by weakly focusing it using a lens with focal length $f_{0}=100$ mm placed before the crystal. In Fig. \ref{fig5}, we present the influence of the angular spectrum of the pump beam on the spatial structure of the polarization states. In Fig. \ref{fig5}(a) and (b), we show the far-field intensity profile of the pump and the two-photon momentum correlation profile for the cases of a collimated pump and a focused pump, respectively. In Fig. \ref{fig5}(b), a single superpixel for the signal photons is spatially correlated with multiple superpixels for the idler photons centered around the conjugate superpixel. In other words, after focusing, the increased angular spectrum of the pump beam substantially widens the two-photon correlation width. In Fig. \ref{fig5}(c) and (e), we depict the results for polarization state tomography for the case of the collimated pump. To illustrate the influence of the transverse momentum of the pump beam on the polarization entanglement of the down-converted photons, we map the concurrence and phase onto pixel coordinates of the joint momentum of signal and idler, which is essentially the transverse momentum of the pump beam $\textbf{\textit{p}}_\text{p}$. Here, the polarization entanglement is found in a few superpixels within the narrower correlation profile, and $\phi$ displays little variation within the narrow angular spectrum of the collimated pump. In Fig. \ref{fig5}(d) and (f), we depict the results for polarization state tomography for the case of the focused pump. The concurrence map indicates the presence of polarization entanglement in all superpixels within the widened correlation profile, with an average concurrence of $0.6930\pm0.0034$. The decrease in concurrence is likely because the focusing of pump broadens the momentum anti-correlated region for a fixed signal superpixel. As a result, detection at each superpixel pair mixes a wider unresolved spread of momentum-dependent $\phi$, thereby reducing the overall polarization entanglement. In contrast to Fig.~\ref{fig5}(e), the phase map illustrates a gradient of $\phi$ dependent on the transverse momentum of the pump beam, which is in good agreement with the theoretical predictions in Fig. 2(a) of \cite{li2023PRA}. To the best of our knowledge, this is the first direct measurement of the cross-influence between the spatial DoF of the pump beam and the polarization DoF of the down-converted photons. Our work opens up the opportunity to control the spatial structure of polarization states by jointly manipulating the polarization and spatial modes of the pump, which has implications in the study of topological structures of quantum light \cite{he2022lsa, shen2024natphot, nape2023aplphot}. For instance, one can pump a paired crystal with a beam with spatially structured polarization and directly shape the correlation between spatial mode and polarization in the nonlocal optical skyrmions\cite{Ornelas2024NP}. 

\section{Conclusion and Outlook}\label{sec4}

In this work, we image the spatially varying polarization entanglement in a two-photon state produced from SPDC using a data-driven camera capable of time-stamping single photons. In contrast to earlier studies that have imposed spatial-polarization structures by modulating a postselected subset of the full SPDC field, our work reveals those that are inherent to the SPDC process. By performing spatially-resolved polarization state tomography, we confirm the generation of an entire class of near-maximally polarization-entangled states with an average concurrence of 0.8303$\pm$0.0004 in the entire spatial profile of the SPDC field. We then present the experimentally characterized spatial-polarization structure. By producing the first complete spatial map of the polarization state generated from SPDC, we illustrate the dependence of polarization entanglement on the transverse momenta of the down-converted photons in the entire spatial profile of the SPDC field. Using a weakly focused pump beam, we, for the first time, experimentally characterize the cross-influence between the polarization DoF of the down-converted photons and the spatial DoF of the pump beam. Although the current Tpx3Cam system has a low overall photon detection efficiency of 8 \% \cite{Vidyapin2023SciRep} and is thus not yet suited for studies involving Bell-type tests of nonlocality \cite{zeitler2022prapp}, its applicability could see further expansion by implementing an image intensifier with higher quantum efficiency and higher gain. For instance, a minimum detection efficiency of $2/3$ is required to close the fair-sampling loophole in Bell tests \cite{Gisin1999PhysLettA}. 

Our results could be important for future work aimed at harnessing SPDC-based entanglement generation for high-dimensional quantum information processing applications \cite{erhard2020natrevphys, friis2019natrevphys}. For instance, time-stamping single photons with a data-driven camera could significantly reduce the acquisition time requirements in polarization entanglement-enabled holography \cite{defienne2021natphys}, potentially enabling fast super-resolution imaging and microscopy \cite{defienne2022natcomm}. Our results on cross-influence between different DoFs could extend the recent demonstration of hiding images in quantum correlations \cite{Verniere2024prl} to hiding and rapidly reconstructing complex phase images in spatial and polarization correlations. On the fundamental side, our work can be extended to explore the spatial-polarization coupling in different entanglement generation schemes. For instance, it is possible to spatially resolve the polarization entanglement produced from post-selecting down-converted photons from a single Type-II nonlinear crystal \cite{Kuklewicz2004pra, zhang2023PRApplied}, thereby further deepening our understanding of the cross-DoF coupling in SPDC. While the present work focuses on the coupling between spatial and polarization DoFs, we believe that, with the help of additional diffraction optics \cite{zhang2021OE}, one can resolve the full spatial-spectral-polarization state and characterize the couplings between all DoFs of light. Furthermore, our work could have important implications for studying and engineering hyperentangled photon states for various applications, such as noise-resilient quantum illumination \cite{Prabhu2021pra}, quantum key distribution with high secure key rates \cite{Wu2017sciencechina, Kim21optica, Nemirovsky-Levy2024opticaquantum}. It may also be possible to build on existing work on exchange phases in Hong-Ou-Mandel interference involving high-dimensional hyperentangled photons \cite{liu2022prl} and spatially resolve such effects for a hyperentangled state.

\begin{acknowledgments}
The authors thank X. Gao, A. D'Errico, B. Braverman, S. Karan, and M. Krenn for fruitful discussions. C. L. acknowledges P. Svihra for helpful advice on data processing. The portion of the work performed at the University of Ottawa was supported by the Canada Research Chairs program under Award 950-231657, the Natural Sciences and Engineering Research Council of Canada under Alliance Consortia Quantum Grant ALLRP 578468 - 22, Discovery Grant RGPIN/2017-06880, and the Canada First Research Excellence Fund Award 072623. In addition, R.W.B. acknowledges support through U.S. National Science Foundation Award No. 2138174 and U.S. Department of Energy Award No. FWP 76295.
\end{acknowledgments}
%\nocite{*}

\bibliography{ms}% Produces the bibliography via BibTeX.

@article{Kwiat1997jmo,
author = {Paul G. Kwiat},
title = {Hyper-entangled states},
journal = {Journal of Modern Optics},
volume = {44},
number = {11-12},
pages = {2173--2184},
year = {1997},
publisher = {Taylor \& Francis},
doi = {10.1080/09500349708231877},
}

@article{Shih1988PRL,
  title = {New Type of {Einstein-Podolsky-Rosen-Bohm} Experiment Using Pairs of Light Quanta Produced by Optical Parametric Down Conversion},
  author = {Shih, Y. H. and Alley, C. O.},
  journal = {Phys. Rev. Lett.},
  volume = {61},
  issue = {26},
  pages = {2921--2924},
  numpages = {0},
  year = {1988},
  month = {Dec},
  publisher = {American Physical Society},
  doi = {10.1103/PhysRevLett.61.2921},
  url = {https://link.aps.org/doi/10.1103/PhysRevLett.61.2921}
}

@article{Harris1967prl,
  title = {Observation of Tunable Optical Parametric Fluorescence},
  author = {Harris, S. E. and Oshman, M. K. and Byer, R. L.},
  journal = {Phys. Rev. Lett.},
  volume = {18},
  issue = {18},
  pages = {732--734},
  numpages = {0},
  year = {1967},
  month = {May},
  publisher = {American Physical Society},
  doi = {10.1103/PhysRevLett.18.732},
  url = {https://link.aps.org/doi/10.1103/PhysRevLett.18.732}
}

@article{Gatti2009prl,
  title = {$X$ Entanglement: The Nonfactorable Spatiotemporal Structure of Biphoton Correlation},
  author = {Gatti, A. and Brambilla, E. and Caspani, L. and Jedrkiewicz, O. and Lugiato, L. A.},
  journal = {Phys. Rev. Lett.},
  volume = {102},
  issue = {22},
  pages = {223601},
  numpages = {4},
  year = {2009},
  month = {Jun},
  publisher = {American Physical Society},
  doi = {10.1103/PhysRevLett.102.223601},
  url = {https://link.aps.org/doi/10.1103/PhysRevLett.102.223601}
}

@article{Baghdasaryan2022pra,
  title = {Generalized description of the spatio-temporal biphoton state in spontaneous parametric down-conversion},
  author = {Baghdasaryan, Baghdasar and Sevilla-Guti\'errez, Carlos and Steinlechner, Fabian and Fritzsche, Stephan},
  journal = {Phys. Rev. A},
  volume = {106},
  issue = {6},
  pages = {063711},
  numpages = {8},
  year = {2022},
  month = {Dec},
  publisher = {American Physical Society},
  doi = {10.1103/PhysRevA.106.063711},
  url = {https://link.aps.org/doi/10.1103/PhysRevA.106.063711}
}

@article{Sevilla-Gutierrez2024pra,
  title = {Spectral properties of transverse Laguerre-Gauss modes in parametric down-conversion},
  author = {Sevilla-Guti\'errez, Carlos and Kaipalath, Varun Raj and Baghdasaryan, Baghdasar and Gr\"afe, Markus and Fritzsche, Stephan and Steinlechner, Fabian},
  journal = {Phys. Rev. A},
  volume = {109},
  issue = {2},
  pages = {023534},
  numpages = {11},
  year = {2024},
  month = {Feb},
  publisher = {American Physical Society},
  doi = {10.1103/PhysRevA.109.023534},
  url = {https://link.aps.org/doi/10.1103/PhysRevA.109.023534}
}

@article{Gisin1999PhysLettA,
  author = {N. Gisin and B. Gisin},
  title = {A local hidden variable model of quantum correlation exploiting the detection loophole},
  journal = {Physics Letters A},
  volume = {260},
  pages = {323-327},
  year = {1999},
  doi = {10.1016/S0375-9601(99)00455-7}
}

@Article{Scarfe2025commphy,
author={Scarfe, Lukas
and Abolhassani, Rojan
and Bouchard, Fr{\'e}d{\'e}ric
and Goldberg, Aaron Z.
and Heshami, Khabat
and Di Colandrea, Francesco
and Karimi, Ebrahim},
title={High-dimensional quantum key distribution with Qubit-like states},
journal={Communications Physics},
year={2025},
month={Nov},
day={25},
volume={8},
number={1},
pages={472},
issn={2399-3650},
doi={10.1038/s42005-025-02376-8},
url={https://doi.org/10.1038/s42005-025-02376-8}
}

@article{Wu2017sciencechina,
  author = {Wu, FangZhou and Yang, GuoJian and Wang, HaiBo and Xiong, Jun and Alzahrani, Faris and Hobiny, Aatef and Deng, FuGuo},
  title = {High-capacity quantum secure direct communication with two-photon six-qubit hyperentangled states},
  journal = {Science China Physics, Mechanics \& Astronomy},
  year = {2017},
  volume = {60},
  pages = {120313},
  doi = {10.1007/s11433-017-9100-9}
}

@article{Nemirovsky-Levy2024opticaquantum,
author = {Liat Nemirovsky-Levy and Uzi Pereg and Mordechai Segev},
journal = {Optica Quantum},
number = {3},
pages = {165--172},
publisher = {Optica Publishing Group},
title = {Increasing quantum communication rates using hyperentangled photonic states},
volume = {2},
month = {Jun},
year = {2024},
url = {https://opg.optica.org/opticaq/abstract.cfm?URI=opticaq-2-3-165},
doi = {10.1364/OPTICAQ.520406},
}

@article{Zhang2020PRA,
  title = {Multidimensional quantum-enhanced target detection via spectrotemporal-correlation measurements},
  author = {Zhang, Yingwen and England, Duncan and Nomerotski, Andrei and Svihra, Peter and Ferrante, Steven and Hockett, Paul and Sussman, Benjamin},
  journal = {Phys. Rev. A},
  volume = {101},
  issue = {5},
  pages = {053808},
  numpages = {9},
  year = {2020},
  month = {May},
  publisher = {American Physical Society},
  doi = {10.1103/PhysRevA.101.053808},
  url = {https://link.aps.org/doi/10.1103/PhysRevA.101.053808}
}

@article{Verniere2024prl,
  title = {Hiding Images in Quantum Correlations},
  author = {Verni\`ere, Chlo\'e and Defienne, Hugo},
  journal = {Phys. Rev. Lett.},
  volume = {133},
  issue = {9},
  pages = {093601},
  numpages = {6},
  year = {2024},
  month = {Aug},
  publisher = {American Physical Society},
  doi = {10.1103/PhysRevLett.133.093601},
  url = {https://link.aps.org/doi/10.1103/PhysRevLett.133.093601}
}

@article{Prabhu2021pra,
  title = {Hyperentanglement-enhanced quantum illumination},
  author = {Prabhu, Ashwith Varadaraj and Suri, Baladitya and Chandrashekar, C. M.},
  journal = {Phys. Rev. A},
  volume = {103},
  issue = {5},
  pages = {052608},
  numpages = {6},
  year = {2021},
  month = {May},
  publisher = {American Physical Society},
  doi = {10.1103/PhysRevA.103.052608},
  url = {https://link.aps.org/doi/10.1103/PhysRevA.103.052608}
}

@article{Kim21optica,
    author = {Jin-Hun Kim and Yosep Kim and Dong-Gil Im and Chung-Hyun Lee and Jin-Woo Chae and Giuliano Scarcelli and Yoon-Ho Kim},
    journal = {Optica},
    number = {12},
    pages = {1524--1531},
    publisher = {Optica Publishing Group},
    title = {Noise-resistant quantum communications using hyperentanglement},
    volume = {8},
    month = {Dec},
    year = {2021},
    url = {https://opg.optica.org/optica/abstract.cfm?URI=optica-8-12-1524},
    doi = {10.1364/OPTICA.442240},
}

@Article{Ornelas2024NP,
author={Ornelas, Pedro
and Nape, Isaac
and de Mello Koch, Robert
and Forbes, Andrew},
title={Non-local skyrmions as topologically resilient quantum entangled states of light},
journal={Nature Photonics},
year={2024},
month={Mar},
day={01},
volume={18},
number={3},
pages={258-266},
issn={1749-4893},
doi={10.1038/s41566-023-01360-4},
url={https://doi.org/10.1038/s41566-023-01360-4}
}

@Article{defienne2022natcomm,
author={Defienne, Hugo
and Cameron, Patrick
and Ndagano, Bienvenu
and Lyons, Ashley
and Reichert, Matthew
and Zhao, Jiuxuan
and Harvey, Andrew R.
and Charbon, Edoardo
and Fleischer, Jason W.
and Faccio, Daniele},
title={Pixel super-resolution with spatially entangled photons},
journal={Nature Communications},
year={2022},
month={Jun},
day={22},
volume={13},
number={1},
pages={3566},
issn={2041-1723},
doi={10.1038/s41467-022-31052-6},
url={https://doi.org/10.1038/s41467-022-31052-6}
}

@article{liu2022prl,
  title = {{Hong-Ou-Mandel} Interference between Two Hyperentangled Photons Enables Observation of Symmetric and Antisymmetric Particle Exchange Phases},
  author = {Liu, Zhi-Feng and Chen, Chao and Xu, Jia-Min and Cheng, Zi-Mo and Ren, Zhi-Cheng and Dong, Bo-Wen and Lou, Yan-Chao and Yang, Yu-Xiang and Xue, Shu-Tian and Liu, Zhi-Hong and Zhu, Wen-Zheng and Wang, Xi-Lin and Wang, Hui-Tian},
  journal = {Phys. Rev. Lett.},
  volume = {129},
  issue = {26},
  pages = {263602},
  numpages = {7},
  year = {2022},
  month = {Dec},
  publisher = {American Physical Society},
  doi = {10.1103/PhysRevLett.129.263602},
  url = {https://link.aps.org/doi/10.1103/PhysRevLett.129.263602}
}

@article{Nomerotski2023joi,
doi = {10.1088/1748-0221/18/01/c01023},
url = {https://doi.org/10.1088/1748-0221/18/01/c01023},
year = {2023},
month = jan,
publisher = {{IOP} Publishing},
volume = {18},
number = {01},
pages = {C01023},
author = {A. Nomerotski and M. Chekhlov and D. Dolzhenko and R. Glazenborg and B. Farella and M. Keach and R. Mahon and D. Orlov and P. Svihra},
title = {Intensified {Tpx3Cam},  a fast data-driven optical camera with nanosecond timing resolution for single photon detection in quantum applications},
journal = {Journal of Instrumentation}
}

@article{zeitler2022prapp,
  title = {Entanglement Verification of Hyperentangled Photon Pairs},
  author = {Zeitler, Christopher K. and Chapman, Joseph C. and Chitambar, Eric and Kwiat, Paul G.},
  journal = {Phys. Rev. Appl.},
  volume = {18},
  issue = {5},
  pages = {054025},
  numpages = {11},
  year = {2022},
  month = {Nov},
  publisher = {American Physical Society},
  doi = {10.1103/PhysRevApplied.18.054025},
  url = {https://link.aps.org/doi/10.1103/PhysRevApplied.18.054025}
}

@Article{shen2024natphot,
author={Shen, Yijie
and Zhang, Qiang
and Shi, Peng
and Du, Luping
and Yuan, Xiaocong
and Zayats, Anatoly V.},
title={Optical skyrmions and other topological quasiparticles of light},
journal={Nature Photonics},
year={2024},
month={Jan},
day={01},
volume={18},
number={1},
pages={15-25},
issn={1749-4893},
doi={10.1038/s41566-023-01325-7},
url={https://doi.org/10.1038/s41566-023-01325-7}
}

@article{HerreraValencia2020Quantum,
  doi = {10.22331/q-2020-12-24-376},
  url = {https://doi.org/10.22331/q-2020-12-24-376},
  title = {High-{D}imensional {P}ixel {E}ntanglement: {E}fficient {G}eneration and {C}ertification},
  author = {Herrera Valencia, Natalia and Srivastav, Vatshal and Pivoluska, Matej and Huber, Marcus and Friis, Nicolai and McCutcheon, Will and Malik, Mehul},
  journal = {{Quantum}},
  issn = {2521-327X},
  publisher = {{Verein zur F{\"{o}}rderung des Open Access Publizierens in den Quantenwissenschaften}},
  volume = {4},
  pages = {376},
  month = dec,
  year = {2020}
}

@article{friis2019natrevphys,
	title = {Entanglement certification from theory to experiment},
	volume = {1},
	rights = {2018 Springer Nature Limited},
	issn = {2522-5820},
	url = {https://www.nature.com/articles/s42254-018-0003-5},
	doi = {10.1038/s42254-018-0003-5},
	pages = {72--87},
	number = {1},
	journal = {Nature Reviews Physics},
	shortjournal = {Nat Rev Phys},
	author = {Friis, Nicolai and Vitagliano, Giuseppe and Malik, Mehul and Huber, Marcus},
	urlyear = {2023},
	year = {2019},
	langid = {english},
	note = {Number: 1
Publisher: Nature Publishing Group},
}

@article{bavaresco2018natphys,
	title = {Measurements in two bases are sufficient for certifying high-dimensional entanglement},
	volume = {14},
	rights = {2018 The Author(s)},
	issn = {1745-2481},
	url = {https://www.nature.com/articles/s41567-018-0203-z},
	doi = {10.1038/s41567-018-0203-z},
	pages = {1032--1037},
	number = {10},
	journal = {Nature Physics},
	shortjournal = {Nature Phys},
	author = {Bavaresco, Jessica and Herrera Valencia, Natalia and Klöckl, Claude and Pivoluska, Matej and Erker, Paul and Friis, Nicolai and Malik, Mehul and Huber, Marcus},
	urlyear = {2023},
	year = {2018},
	langid = {english},
	note = {Number: 10
Publisher: Nature Publishing Group},
}

@article{erker2017quantum,
	title = {Quantifying high dimensional entanglement with two mutually unbiased bases},
	volume = {1},
	url = {https://quantum-journal.org/papers/q-2017-07-28-22/},
	doi = {10.22331/q-2017-07-28-22},
	pages = {22},
	journal = {Quantum},
	author = {Erker, Paul and Krenn, Mario and Huber, Marcus},
	urlyear = {2023},
	year = {2017},
	langid = {british},
	note = {Publisher: Verein zur Förderung des Open Access Publizierens in den Quantenwissenschaften},
}

@article{courme2023optlett,
	title = {Quantifying high-dimensional spatial entanglement with a single-photon-sensitive time-stamping camera},
	volume = {48},
	rights = {© 2023 Optica Publishing Group},
	issn = {1539-4794},
	url = {https://opg.optica.org/ol/abstract.cfm?uri=ol-48-13-3439},
	doi = {10.1364/OL.487182},
	pages = {3439--3442},
	number = {13},
	journal = {Optics Letters},
	shortjournal = {Opt. Lett., {OL}},
	author = {Courme, Baptiste and Vernière, Chloé and Svihra, Peter and Gigan, Sylvain and Nomerotski, Andrei and Defienne, Hugo},
	urlyear = {2023},
	year = {2023},
	note = {Publisher: Optica Publishing Group},
}

@article{law2004prl,
	title = {Analysis and Interpretation of High Transverse Entanglement in Optical Parametric Down Conversion},
	volume = {92},
	url = {https://link.aps.org/doi/10.1103/PhysRevLett.92.127903},
	doi = {10.1103/PhysRevLett.92.127903},
	pages = {127903},
	number = {12},
	journal = {Physical Review Letters},
	shortjournal = {Phys. Rev. Lett.},
	author = {Law, C. K. and Eberly, J. H.},
	urlyear = {2023},
	year = {2004},
	note = {Publisher: American Physical Society},
}

@article{ndagano2020npjqi,
	title = {Imaging and certifying high-dimensional entanglement with a single-photon avalanche diode camera},
	volume = {6},
	rights = {2020 The Author(s)},
	issn = {2056-6387},
	url = {https://www.nature.com/articles/s41534-020-00324-8},
	doi = {10.1038/s41534-020-00324-8},
	pages = {1--8},
	number = {1},
	journal = {npj Quantum Information},
	shortjournal = {npj Quantum Inf},
	author = {Ndagano, Bienvenu and Defienne, Hugo and Lyons, Ashley and Starshynov, Ilya and Villa, Federica and Tisa, Simone and Faccio, Daniele},
	urlyear = {2023},
	year = {2020},
	langid = {english},
	note = {Number: 1
Publisher: Nature Publishing Group},
}

@article{Fisher2016joi,
    title={TimepixCam: a fast optical imager with time-stamping},
    author={Fisher-Levine, M. and Nomerotski, A.},
    journal={Journal of Instrumentation},
    volume={11},
    number={03},
    pages={C03016},
    year={2016},
    publisher={IOP Publishing}
}

@article{defienne2021natphys,
	title = {Polarization entanglement-enabled quantum holography},
	volume = {17},
	rights = {2021 The Author(s), under exclusive licence to Springer Nature Limited},
	issn = {1745-2481},
	url = {https://www.nature.com/articles/s41567-020-01156-1},
	doi = {10.1038/s41567-020-01156-1},
	pages = {591--597},
	number = {5},
	journal = {Nature Physics},
	shortjournal = {Nat. Phys.},
	author = {Defienne, Hugo and Ndagano, Bienvenu and Lyons, Ashley and Faccio, Daniele},
	urlyear = {2023},
	year = {2021},
	langid = {english},
	note = {Number: 5
Publisher: Nature Publishing Group},
}

@article{barreiro2005prl,
	title = {Generation of Hyperentangled Photon Pairs},
	volume = {95},
	url = {https://link.aps.org/doi/10.1103/PhysRevLett.95.260501},
	doi = {10.1103/PhysRevLett.95.260501},
	pages = {260501},
	number = {26},
	journal = {Physical Review Letters},
	shortjournal = {Phys. Rev. Lett.},
	author = {Barreiro, Julio T. and Langford, Nathan K. and Peters, Nicholas A. and Kwiat, Paul G.},
	urlyear = {2023},
	year = {2005},
	note = {Publisher: American Physical Society},
}

@article{erhard2020natrevphys,
	title = {Advances in high-dimensional quantum entanglement},
	volume = {2},
	rights = {2020 Springer Nature Limited},
	issn = {2522-5820},
	url = {https://www.nature.com/articles/s42254-020-0193-5},
	doi = {10.1038/s42254-020-0193-5},
	pages = {365--381},
	number = {7},
	journal = {Nature Reviews Physics},
	shortjournal = {Nat Rev Phys},
	author = {Erhard, Manuel and Krenn, Mario and Zeilinger, Anton},
	urlyear = {2023},
	year = {2020},
	langid = {english},
	note = {Number: 7
Publisher: Nature Publishing Group},
}

@article{nape2023aplphot,
	title = {Quantum structured light in high dimensions},
	volume = {8},
	issn = {2378-0967},
	url = {https://doi.org/10.1063/5.0138224},
	doi = {10.1063/5.0138224},
	pages = {051101},
	number = {5},
	journal = {{APL} Photonics},
	shortjournal = {{APL} Photonics},
	author = {Nape, Isaac and Sephton, Bereneice and Ornelas, Pedro and Moodley, Chane and Forbes, Andrew},
	urlyear = {2023},
	year = {2023},
}

@article{he2022lsa,
	title = {Towards higher-dimensional structured light},
	volume = {11},
	rights = {2022 The Author(s)},
	issn = {2047-7538},
	url = {https://www.nature.com/articles/s41377-022-00897-3},
	doi = {10.1038/s41377-022-00897-3},
	pages = {205},
	number = {1},
	journal = {Light: Science \& Applications},
	shortjournal = {Light Sci Appl},
	author = {He, Chao and Shen, Yijie and Forbes, Andrew},
	urlyear = {2023},
	year = {2022},
	langid = {english},
	note = {Number: 1
Publisher: Nature Publishing Group},
}

@article{wootters1998prl,
	title = {Entanglement of Formation of an Arbitrary State of Two Qubits},
	volume = {80},
	url = {https://link.aps.org/doi/10.1103/PhysRevLett.80.2245},
	doi = {10.1103/PhysRevLett.80.2245},
	pages = {2245--2248},
	number = {10},
	journal = {Physical Review Letters},
	shortjournal = {Phys. Rev. Lett.},
	author = {Wootters, William K.},
	urlyear = {2023},
	year = {1998},
	note = {Publisher: American Physical Society},
}

@book{Nielsen_Chuang_2010, place={Cambridge}, title={Quantum Computation and Quantum Information: 10th Anniversary Edition}, publisher={Cambridge University Press}, author={Nielsen, Michael A. and Chuang, Isaac L.}, year={2010}}

@article{wootters2001qic,
author = {Wootters, William K.},
title = {Entanglement of formation and concurrence},
year = {2001},
issue_date = {January 2001},
publisher = {Rinton Press, Incorporated},
address = {Paramus, NJ},
volume = {1},
number = {1},
issn = {1533-7146},
journal = {Quantum Info. Comput.},
month = {jan},
pages = {27–44},
numpages = {18}
}

@Article{Vidyapin2023SciRep,
author={Vidyapin, Victor
and Zhang, Yingwen
and England, Duncan
and Sussman, Benjamin},
title={Characterisation of a single photon event camera for quantum imaging},
journal={Scientific Reports},
year={2023},
month={Jan},
day={18},
volume={13},
number={1},
pages={1009},
issn={2045-2322},
doi={10.1038/s41598-023-27842-7},
url={https://doi.org/10.1038/s41598-023-27842-7}
}

@article{Gao2024PRL,
  title = {Full Spatial Characterization of Entangled Structured Photons},
  author = {Gao, Xiaoqin and Zhang, Yingwen and D'Errico, Alessio and Sit, Alicia and Heshami, Khabat and Karimi, Ebrahim},
  journal = {Phys. Rev. Lett.},
  volume = {132},
  issue = {6},
  pages = {063802},
  numpages = {7},
  year = {2024},
  month = {Feb},
  publisher = {American Physical Society},
  doi = {10.1103/PhysRevLett.132.063802},
  url = {https://link.aps.org/doi/10.1103/PhysRevLett.132.063802}
}

@article{james2001PRA,
	title = {Measurement of qubits},
	volume = {64},
	url = {https://link.aps.org/doi/10.1103/PhysRevA.64.052312},
	doi = {10.1103/PhysRevA.64.052312},
	pages = {052312},
	number = {5},
	journal = {Physical Review A},
	shortjournal = {Phys. Rev. A},
	author = {James, Daniel F. V. and Kwiat, Paul G. and Munro, William J. and White, Andrew G.},
	urlyear = {2023},
	year = {2001},
	note = {Publisher: American Physical Society},
}

@article{kwiat1999pra,
	title = {Ultrabright source of polarization-entangled photons},
	volume = {60},
	url = {https://link.aps.org/doi/10.1103/PhysRevA.60.R773},
	doi = {10.1103/PhysRevA.60.R773},
	pages = {R773--R776},
	number = {2},
	journal = {Physical Review A},
	shortjournal = {Phys. Rev. A},
	author = {Kwiat, Paul G. and Waks, Edo and White, Andrew G. and Appelbaum, Ian and Eberhard, Philippe H.},
	urlyear = {2023},
	year = {1999},
	note = {Publisher: American Physical Society},
}

@article{altepeter2005opex,
	title = {Phase-compensated ultra-bright source of entangled photons},
	volume = {13},
	rights = {© 2005 Optical Society of America},
	issn = {1094-4087},
	url = {https://opg.optica.org/oe/abstract.cfm?uri=oe-13-22-8951},
	doi = {10.1364/OPEX.13.008951},
	pages = {8951--8959},
	number = {22},
	journal = {Optics Express},
	shortjournal = {Opt. Express, {OE}},
	author = {Altepeter, J. B. and Jeffrey, E. R. and Kwiat, P. G.},
	urlyear = {2023},
	year = {2005},
	note = {Publisher: Optica Publishing Group},
}

@article{li2023PRA,
	title = {Experimental generation of polarization entanglement from spontaneous parametric down-conversion pumped by spatiotemporally highly incoherent light},
	volume = {107},
	url = {https://link.aps.org/doi/10.1103/PhysRevA.107.L041701},
	doi = {10.1103/PhysRevA.107.L041701},
	pages = {L041701},
	number = {4},
	journal = {Physical Review A},
	shortjournal = {Phys. Rev. A},
	author = {Li, Cheng and Braverman, Boris and Kulkarni, Girish and Boyd, Robert W.},
	urlyear = {2023},
	year = {2023},
	note = {Publisher: American Physical Society},
}

@article{burnham1970PRL,
	title = {Observation of Simultaneity in Parametric Production of Optical Photon Pairs},
	volume = {25},
	url = {https://link.aps.org/doi/10.1103/PhysRevLett.25.84},
	doi = {10.1103/PhysRevLett.25.84},
	pages = {84--87},
	number = {2},
	journal = {Physical Review Letters},
	shortjournal = {Phys. Rev. Lett.},
	author = {Burnham, David C. and Weinberg, Donald L.},
	urlyear = {2023},
	year = {1970},
	note = {Publisher: American Physical Society},
}

@article{nambu2002PRA,
	title = {Generation of polarization-entangled photon pairs in a cascade of two type-I crystals pumped by femtosecond pulses},
	volume = {66},
	url = {https://link.aps.org/doi/10.1103/PhysRevA.66.033816},
	doi = {10.1103/PhysRevA.66.033816},
	pages = {033816},
	number = {3},
	journal = {Physical Review A},
	shortjournal = {Phys. Rev. A},
	author = {Nambu, Yoshihiro and Usami, Koji and Tsuda, Yoshiyuki and Matsumoto, Keiji and Nakamura, Kazuo},
	urlyear = {2023},
	year = {2002},
	note = {Publisher: American Physical Society},
}

@article{Vallone2010PRA,
  title = {Six-qubit two-photon hyperentangled cluster states: Characterization and application to quantum computation},
  author = {Vallone, Giuseppe and Donati, Gaia and Ceccarelli, Raino and Mataloni, Paolo},
  journal = {Phys. Rev. A},
  volume = {81},
  issue = {5},
  pages = {052301},
  numpages = {10},
  year = {2010},
  month = {May},
  publisher = {American Physical Society},
  doi = {10.1103/PhysRevA.81.052301},
  url = {https://link.aps.org/doi/10.1103/PhysRevA.81.052301}
}

@article{Ciampini2016Light,
author={Ciampini, Mario Arnolfo
and Orieux, Adeline
and Paesani, Stefano
and Sciarrino, Fabio
and Corrielli, Giacomo
and Crespi, Andrea
and Ramponi, Roberta
and Osellame, Roberto
and Mataloni, Paolo},
title={Path-polarization hyperentangled and cluster states of photons on a chip},
journal={Light: Science {\&} Applications},
year={2016},
month={Apr},
day={01},
volume={5},
number={4},
pages={e16064-e16064},
issn={2047-7538},
doi={10.1038/lsa.2016.64},
url={https://doi.org/10.1038/lsa.2016.64}
}

@article{ianzano2020SciRep,
	title = {Fast camera spatial characterization of photonic polarization entanglement},
	volume = {10},
	rights = {2020 The Author(s)},
	issn = {2045-2322},
	url = {https://www.nature.com/articles/s41598-020-62020-z},
	doi = {10.1038/s41598-020-62020-z},
	pages = {6181},
	number = {1},
	journal = {Scientific Reports},
	shortjournal = {Sci Rep},
	author = {Ianzano, Christopher and Svihra, Peter and Flament, Mael and Hardy, Andrew and Cui, Guodong and Nomerotski, Andrei and Figueroa, Eden},
	urlyear = {2023},
	year = {2020},
	langid = {english},
	note = {Number: 1
Publisher: Nature Publishing Group},
}

@article{zhang2021OE,
	title = {High speed imaging of spectral-temporal correlations in {Hong-Ou-Mandel} interference},
	volume = {29},
	rights = {© 2021 Optical Society of America},
	issn = {1094-4087},
	url = {https://opg.optica.org/oe/abstract.cfm?uri=oe-29-18-28217},
	doi = {10.1364/OE.432191},
	pages = {28217--28227},
	number = {18},
	journal = {Optics Express},
	shortjournal = {Opt. Express, {OE}},
	author = {Zhang, Yingwen and England, Duncan and Nomerotski, Andrei and Sussman, Benjamin},
	urlyear = {2023},
	year = {2021},
	note = {Publisher: Optica Publishing Group},
}

@article{Karan_2020,
doi = {10.1088/2040-8986/ab89e4},
url = {https://doi.org/10.1088/2040-8986/ab89e4},
year = {2020},
month = {jun},
publisher = {IOP Publishing},
volume = {22},
number = {8},
pages = {083501},
author = {Karan, Suman and Aarav, Shaurya and Bharadhwaj, Homanga and Taneja, Lavanya and De, Arinjoy and Kulkarni, Girish and Meher, Nilakantha and Jha, Anand K},
title = {Phase matching in $\beta$-barium borate crystals for spontaneous parametric down-conversion},
journal = {Journal of Optics}
}

@article{Brida2007pra,
  title = {Generation of different Bell states within the spontaneous parametric down-conversion phase-matching bandwidth},
  author = {Brida, Giorgio and Chekhova, Maria and Genovese, Marco and Krivitsky, Leonid},
  journal = {Phys. Rev. A},
  volume = {76},
  issue = {5},
  pages = {053807},
  numpages = {10},
  year = {2007},
  month = {Nov},
  publisher = {American Physical Society},
  doi = {10.1103/PhysRevA.76.053807},
  url = {https://link.aps.org/doi/10.1103/PhysRevA.76.053807}
}

@article{reid1989PRA,
	title = {Demonstration of the {Einstein–Podolsky–Rosen} paradox using nondegenerate parametric amplification},
	volume = {40},
	url = {https://link.aps.org/doi/10.1103/PhysRevA.40.913},
	doi = {10.1103/PhysRevA.40.913},
	pages = {913--923},
	number = {2},
	journal = {Physical Review A},
	shortjournal = {Phys. Rev. A},
	author = {Reid, M. D.},
	urlyear = {2023},
	year = {1989},
	note = {Publisher: American Physical Society},
}

@article{monken1998pra,
	title = {Transfer of angular spectrum and image formation in spontaneous parametric down-conversion},
	volume = {57},
	url = {https://link.aps.org/doi/10.1103/PhysRevA.57.3123},
	doi = {10.1103/PhysRevA.57.3123},
	pages = {3123--3126},
	number = {4},
	journal = {Physical Review A},
	shortjournal = {Phys. Rev. A},
	author = {Monken, C. H. and Ribeiro, P. H. Souto and Pádua, S.},
	urlyear = {2023},
	year = {1998},
	note = {Publisher: American Physical Society},
}

@article{bhattacharjee2022njp,
  title={Measurement of two-photon position--momentum Einstein--Podolsky--Rosen correlations through single-photon intensity measurements},
  author={Bhattacharjee, Abhinandan and Meher, Nilakantha and Jha, Anand K},
  journal={New Journal of Physics},
  volume={24},
  number={5},
  pages={053033},
  year={2022},
  publisher={IOP Publishing}
}

@article{nomerotski2019nuclear,
	title = {Imaging and time stamping of photons with nanosecond resolution in Timepix based optical cameras},
	volume = {937},
	issn = {0168-9002},
	url = {https://www.sciencedirect.com/science/article/pii/S0168900219306667},
	doi = {10.1016/j.nima.2019.05.034},
	pages = {26--30},
	journal = {Nuclear Instruments and Methods in Physics Research Section A: Accelerators, Spectrometers, Detectors and Associated Equipment},
	shortjournal = {Nuclear Instruments and Methods in Physics Research Section A: Accelerators, Spectrometers, Detectors and Associated Equipment},
	author = {Nomerotski, Andrei},
	urlyear = {2023},
	year = {2019},
}

@article{Kuklewicz2004pra,
  title = {High-flux source of polarization-entangled photons from a periodically poled {KTiOPO}$_4$ parametric down-converter},
  author = {Kuklewicz, Christopher E. and Fiorentino, Marco and Messin, Ga\'etan and Wong, Franco N. C. and Shapiro, Jeffrey H.},
  journal = {Phys. Rev. A},
  volume = {69},
  issue = {1},
  pages = {013807},
  numpages = {5},
  year = {2004},
  month = {Jan},
  publisher = {American Physical Society},
  doi = {10.1103/PhysRevA.69.013807},
  url = {https://link.aps.org/doi/10.1103/PhysRevA.69.013807}
}

@article{zhang2023PRApplied,
	title = {Polarization Entanglement from Parametric Down-conversion with an {LED} Pump},
	volume = {19},
	url = {https://link.aps.org/doi/10.1103/PhysRevApplied.19.054079},
	doi = {10.1103/PhysRevApplied.19.054079},
	pages = {054079},
	number = {5},
	journal = {Physical Review Applied},
	shortjournal = {Phys. Rev. Appl.},
	author = {Zhang, Wuhong and Xu, Diefei and Chen, Lixiang},
	urlyear = {2023},
	year = {2023},
	note = {Publisher: American Physical Society},
}
\end{document}